\documentclass[aps,twocolumn,floatfix]{revtex4}

\usepackage{amsmath,amsthm,amsfonts,amssymb,times,bbm,graphicx}

\newtheorem{theorem}{Theorem}
\newtheorem{proposition}[theorem]{Proposition}
\newtheorem{lemma}[theorem]{Lemma}
\newtheorem{corollary}[theorem]{Corollary}
\newtheorem{definition}[theorem]{Definition}
\newtheorem{conjecture}[theorem]{Conjecture}

\def\Symp#1,#2,#3,#4.{\left[\left(\begin{array}{c}#1\\#2\end{array}\right),\left(\begin{array}{c}#3\\#4\end{array}\right)\right]}
\def\Vec#1,#2.{\left(\!\begin{array}{c}#1\\#2\end{array}\!\right)}
\def\tuple#1.{\langle#1\rangle}
\def\ket#1.{|#1\rangle}
\def\bra#1.{\langle#1|}
\def\braket#1,#2.{\langle#1|#2\rangle}

\def\ZZ{\mathbbm{Z}}
\def\RR{\mathbbm{R}}
\def\CC{\mathbbm{C}}
\def\FF{\mathbbm{F}}
\def\NN{\mathbbm{N}}

\def\H{\mathcal{H}}
\def\D{\mathcal{D}}
\def\Id{\mathbbm{1}}
\def\spacedot{\,\cdot\,}

\DeclareMathOperator{\tr}{tr}
\DeclareMathOperator{\Tr}{Tr}
\DeclareMathOperator{\Sp}{Sp}

\DeclareMathOperator{\rank}{rank}

\begin{document}

\title{Evenly distributed unitaries: on the structure of unitary designs}

\author{D.\ Gross, K.\ Audenaert, and J.\ Eisert}

\affiliation{
	Institute for Mathematical Sciences, Imperial College London, 
	Princes Gate, London
	SW7 2PE, UK
} 
\affiliation{
	QOLS, Blackett Laboratory, Imperial College London, Prince Consort Road, 
	London SW7 2BW, UK
}

\email{david.gross@imperial.ac.uk}

\date{\today}

\begin{abstract} 
We clarify the mathematical structure underlying \emph{unitary
$t$-designs}. These are sets of unitary matrices, evenly distributed
in the sense that the average of any $t$-th order polynomial over the
design equals the average over the entire unitary group. We present a
simple necessary and sufficient criterion for deciding if a set of
matrices constitutes a design. Lower bounds for the number of elements
of $2$-designs are derived. We show how to turn mutually unbiased
bases into approximate 2-designs whose cardinality is optimal in
leading order. Designs of higher order are discussed and an example of
a unitary 5-design is presented. We comment on the relation between
unitary and spherical designs and outline methods for finding designs
numerically or by searching character tables of finite groups.
Further, we sketch connections to problems in linear optics
and questions regarding typical entanglement.
\end{abstract}


\maketitle

\section{Introduction}

Before introducing the notion of a unitary design
it is worthwhile to look at the analogue
structure on spheres in $\RR^n$. Imagine one is interested in the
average value of a real function $f$ defined on an $n$-dimensional
real sphere $\mathcal{S}^n$. That value might be hard to compute in
general so it could be sensible to estimate it by averaging over a
finite set of unit vectors
$\mathcal{D}=\{\ket\psi_1.,\dots,\ket\psi_K.\}$. Of course, for any
such finite set, there are functions whose true average value deviates
arbitrarily much from the one approximated by summing over
$\mathcal{D}$; but the more points the test set includes and the more
``even'' these vectors are distributed, the more ``exotic'' such functions
have to be. The following notion aims to quantitatively capture the
quality of a set of points for these purposes: a finite subset
$\mathcal{D}$ of $S^n$ is called a \emph{spherical $t$-design} if the
average of every $t$-th order polynomial $p$ over $\mathcal{S}^n$
equals $p$'s average taken over $\mathcal{D}$.  A large body of
literature has been devoted to the construction and exploration of
designs. Many of the relevant references can be found in the
accessible article Ref.\ \cite{koenig}.

One can adapt the definition of spherical designs to complex vector
spaces (simply by substituting the real sphere by the set of complex unit
vectors) with obvious applications in quantum mechanics.
In the context of quantum
information theory, $2$-designs appeared in Refs.\
\cite{sics,klappenecker,zauner,iblisdir,scott,hayashi}, to name a few.
Two prominent examples of complex spherical 2-designs are here known by
the names of \emph{mutually unbiased bases}
\cite{mubs,koenig,klappenecker} and \emph{symmetric
informationally complete POVMs} \cite{sics,koenig,klappenecker}
respectively.

Quite recently, Dankert et al.\ introduced the notion of a
\emph{unitary $t$-design} by replacing the 
real sphere $\mathcal{S}^n$
by the set of unitary matrices $U(d)$ in the definition of spherical
designs \cite{dankertThesis,dankert}. 
Averages are here to be taken with respect to the Haar
measure.  In the sense made precise above, the theory of unitary
designs thus aims to identify finite nets of unitaries, which cover
the entire group as tightly as possible.  

Such nets are interesting for various reasons. Abstractly, unitary
designs can serve as testbeds for examining conjectures concerning the
unitary group: their even distribution here means that they ``cover
many complementary aspects'' of $U(d)$. Also, spherical designs
naturally appear in optimal solutions to several physical problems,
ranging from quantum state tomography and optimal estimation using
finite ensembles to quantum key distribution
\cite{sics,iblisdir,hayashi,Fuchs} -- it is sensible to assume that
similar applications for their unitary counterpart can be found.  
More
concretely, unitary designs have been applied to 
quantum process estimation and fidelity estimation of channels
using random states \cite{dankertThesis,dankert}; quantum cryptography
\cite{chau} and data hiding protocols \cite{dataHiding}.
Naturally, designs can be used to estimate
Haar averages using classical computers. 
For the case of
averages of polynomial functions, the results are guaranteed to be
correct (there are, however, other methods for tackling this specific
problem; see Appendix \ref{sec:generalTwirling} and Ref.\
\cite{geza}). For non-polynomial functions one obtains at least an
educated guess.  
Going beyond finite-dimensional quantum systems, Haar
averages over $U(d)$ appear in the context of
energy-preserving transformations of $d$ bosonic modes.
Such transformations are notably relevant as 
passive linear optical transformations of states of light
modes \cite{alessio,GBook}.  
Lastly, we believe the problem to be of inherent 
geometrical interest.

As most of the present work is concerned with unitary $2$-designs, we
now state the precise definition for this special case (see, however,
Section \ref{sec:higherOrder}):

\begin{definition}[Unitary design \cite{dankert,dankertThesis}]
	\label{def:design}
	A set $\mathcal{D}=\{U_k\}_{k=1,\dots, K}$ of unitary matrices on
	$\H=\CC^d$ is a \emph{unitary 2-design} if it fulfills the
	equivalent conditions:
	\begin{enumerate}
		\item\emph{(Averages)}
		Let $p$ be a polynomial in $2d^2$ variables. We can conceive $p$ as a
		function on $U(d)$ by evaluating it on the matrix elements and their
		complex conjugates of a given matrix: $p(U):=p({U^i}_j, \bar
		{U^i}_j)$. One now demands that for any $p$ which is homogeneous of
		degree two in each variable, the relation 
		\begin{equation}\label{eqn:design}
			\frac1K \sum_{U_k\in\mathcal{D}} p(U_k) = \int_{U(d)} p(U) dU
		\end{equation}
		be fulfilled.

		\item\emph{(Twirling of states)}
		For all $\rho \in \mathcal{B}(\mathcal{H}\otimes\mathcal{H})$
		\begin{eqnarray}\label{eqn:uutwirling}
			&&\frac1K \sum_{U_k\in \mathcal{D}} 
			(U_k\otimes U_k) \,\rho\, (U_k\otimes
			U_k)^\dagger \\
			&=& \int_{U(d)} (U\otimes U) \,\rho\, (U\otimes U)^\dagger
			dU. \label{eqn:contuutwirling}
		\end{eqnarray}

		\item
		\emph{(Twirling of channels)}
		For any quantum channel $\Lambda$
		\begin{eqnarray}\label{eqn:depol}
			&&\frac1K \sum_{U_k\in\mathcal{D}} U_k^\dagger 
			\Lambda(U_k \rho U_k^\dagger) U_k \\
			&=& \int_{U(d)} U^\dagger \Lambda(U \rho U^\dagger) U dU. \nonumber
		\end{eqnarray}
	\end{enumerate}
\end{definition}

The problem has a long history, which is formulated mostly in the
second of the three equivalent guises listed above. The ``twirling''
operation originates from invariant theory (where it is sometimes
called ``transfer homomorphism'') and has, to our knowledge, first
been introduced to quantum information theory in Ref.\
\cite{wernerOld}, giving rise to the concept of a ``Werner state''. 
Later, it was noted that in $d=2$ (i.e., for single qubits), it
suffices to average over a finite set of unitaries
\cite{mixedStateEnt}. A construction for general dimensions --
employing non-evenly weighted unitaries -- appeared in Ref.\
\cite{duer}.  DiVincenzo et al.\ \cite{dataHiding} realized that the
\emph{Clifford group} \cite{gottesman,clifford} for qubit systems
exhibits the property given in Eq.\ (\ref{eqn:uutwirling}); a fact
which was later generalized to systems of prime-power dimensions by
Chau \cite{chau}.  Similar ideas appeared in Ref.\ \cite{duer2}. A
first concise treatment was given in a master thesis by Dankert
\cite{dankertThesis} (where the term of a \emph{unitary $t$-design}
has been coined) and in a later paper by Dankert et al.\
\cite{dankert}. In these publications, the equivalence of the criteria
in Definition \ref{def:design} has been made explicit and the question
of how to efficiently implement the unitaries of certain designs was
addressed.

Despite the large amount of interest paid to the problem, the
following natural questions have been left open and will partly be
answered in this paper:
\begin{enumerate}
	\item
	\emph{In which dimensions do unitary $2$-designs exist and when can
	they be explicitly constructed?}  While we do
	not have a general answer to this question, a host of examples is
	provided in Sections \ref{sec:groupDesigns} and
	\ref{sec:cliffordDesigns}. [Note added in revised version: After
	this article had been submitted, A.\ Scott made us aware of Ref.
	\cite{seymour}. This extremely general paper proves -- among other
	things -- the existence of unitary designs for every $t$ and $d$ (it
	does not provide an explicitly way for constructing the designs).
	Thus, the question posed above can partly be answered
	affirmatively.]

	\item
	\emph{What is the minimal number of elements needed for a
	$2$-design?}
	See Section \ref{sec:bounds} for a lower bound, which we conjecture
	to be tight in leading order.

	\item
	\emph{Is there an easy criterion to decide whether a given set of
	matrices constitutes a design?} This question is answered
	affirmatively in Section \ref{sec:framePotential}. We transfer the
	concept of a \emph{frame potential} \cite{benedetto,koenig} from
	spherical to unitary designs. The frame potential is a simple
	polynomial expression in the matrix elements, which is minimized
	exactly for designs.  This criterion even allows for numerical
	searches in spaces of small dimensions.

	\item
	\emph{Can one find designs among matrix groups?} Section
	\ref{sec:groupDesigns} treats this special case. It turns out that
	the theory is especially clear when one restricts attention to
	groups. The frame potential will be re-interpreted in terms of
	basic character theory.

	\item
	\emph{Is it possible to explicitly construct approximate unitary
	designs?} In Section \ref{sec:approximations} we give an explicit
	construction for turning mutually unbiased bases (MUBs) into unitary
	matrices which asymptotically approximate 2-designs. More precisely,
	the prescription yields a set of unitaries for every prime-power
	dimension $d$. These sets approximate 2-designs as $d\to\infty$.
	Also, the cardinality of such \emph{asymptotic designs} is of the same
	order as the lower bound derived earlier.

	\item 
	The set of operators $\mathcal{B}(\CC^{d})$ on $\CC^d$ form a
	$d^2$-dimensional Hilbert space. \emph{What is the connection
	between the unitary designs in $U(d)$ and spherical designs in
	$\CC^{d^2}$?} The relation can be made rather explicit in terms of
	the Jamio\l kowski isomorphism and the frame potential. 
	Both
	spherical and unitary designs correspond to minima 
	of the potential
	-- yet under different constraints. This statement is made precise
	in Section \ref{sec:framePotential}.

	\item
	\emph{What about more general concepts such as $t$-designs for $t>2$
	or substituting $U(d)$ by other groups?} We will discuss this
	general scenario in Section \ref{sec:higherOrder} and present an
	example of a qubit 5-design.
\end{enumerate}

\begin{figure}
  \centering
	(a)\hspace{-1em}
  \includegraphics[scale=.5]{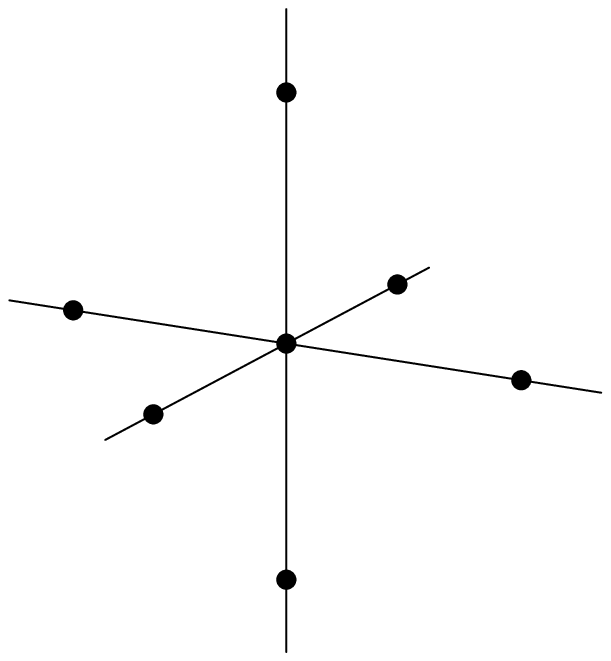}
  \qquad
	(b)\hspace{-1em}
  \includegraphics[scale=.5]{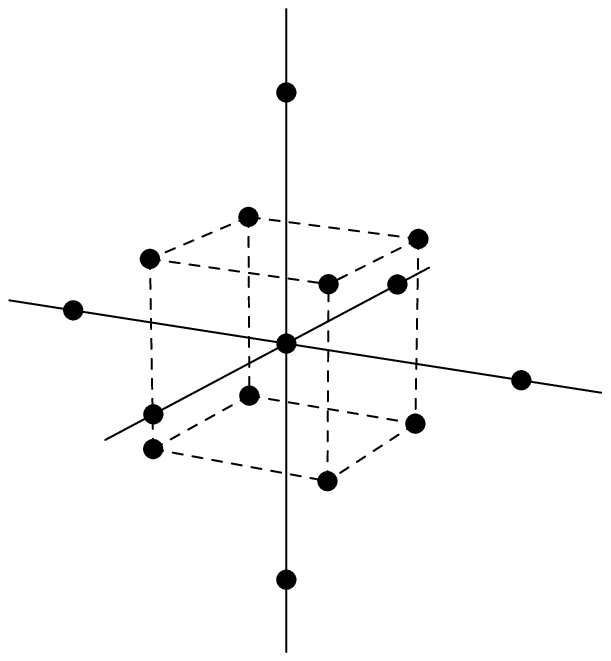}
  \caption{
    \label{fig:qubitDesign}
    Visualization of the 12-element Clifford 2-design described in
    Section \ref{sec:cliffordDesigns}. As up to phases $SU(2)\simeq
    SO(3)$,
    every qubit unitary corresponds to a three-dimensional
    rotation.  The group $SO(3)$, in turn, can be pictured as a ball
    with radius $\pi$, where antipodes on the boundary are identified.
    This is done by associating to every rotation by an angle
    $\phi\in[0,\pi]$ about the unit-vector $\hat n$ the point $\phi\,
    \hat n\in \RR^3$. Figure (a) shows the four Pauli matrices $\Id,
    \sigma_x,\sigma_y,\sigma_z$ in this representation. The
    non-trivial Pauli operations lie on the boundary of the ball and
    hence appear twice: $\sigma_x$, e.g., at $\pm (\pi,0,0)^T$. Adding
    eight further Clifford operations, which correspond to the
    vertices $2\pi/\sqrt{27} (\pm 1, \pm 1, \pm 1)^T$ of a cube, we
    arrive at the 2-design pictured in Figure (b).
  }
\end{figure}

\section{General theory}
\label{sec:general}

\subsection{Preliminaries}
\label{sec:generalPreliminaries}

In this section we are going to derive a simple criterion for
identifying 2-designs as well as lower bounds for the number of
elements $K$ they need to contain. Before stating these results, let
us shortly recall some general facts about \emph{twirling channels}
and \emph{completely positive maps} which will be needed in the
sequel.

Let $\{U_g\}_{g\in G}$ be a unitary representation of some group $G$
on a Hilbert space $\H$.  The \emph{twirling channel}
induced by $G$ and the representation $U_g$ is
\begin{equation}\label{eqn:generaltwirling}
	T(A)=\int_g U_g A U_g^\dagger\,dg,
\end{equation}
where $dg$ stands for the Haar measure of the group $G$.
Denote the projection operators onto the irreducible subspaces of
$\{U_g\}_g$ by $\{P_i\}_i$. For simplicity we assume that the
representation is a direct sum of \emph{inequivalent} irreducible
ones (see Appendix \ref{sec:generalTwirling} for the general case).
By Schur's Lemma $T(A)=A$ if and only if $A$ is a linear combination
of the $P_i$'s. Indeed, setting $P_i':=P_i/{\tr P_i}$, one easily
checks that \begin{equation}\label{eqn:twirlingProjections}
	T(A) = \sum_i \tr(P_i' A) P_i.
\end{equation}

Setting $\H=\CC^d\otimes \CC^d$, $G=U(d)$ represented as $U\mapsto
U\otimes U$, we arrive at the \emph{$UU$-twirling} channel $T_{UU}$
defined in Eq.\  (\ref{eqn:contuutwirling}), which has played a
prominent role in quantum information theory. In order to
identify the irreducible subspaces, define the \emph{flip operator}
$\FF$ which acts by permuting the tensor factors: 
$\FF \ket i. \otimes \ket j. = \ket j.\otimes \ket i.$.  
Its eigenspaces are the sets of \emph{symmetric} and
\emph{anti-symmetric} vectors respectively. The projection operators
onto these spaces will be denoted by $P_S=(\Id + \FF)/2$ and
$P_A=(\Id-\FF)/2$. We have that $\dim P_S = d(d+1)/2$ and $\dim P_A = 
d(d-1)/2$.

Moving on, we recall the well-known correspondence between completely
positive maps (\emph{cp maps}) sending
$\mathcal{B}(\CC^d)\to\mathcal{B}(\CC^d)$ and states on
$\CC^d\otimes\CC^d$. Let $\Lambda$ be such a map.  Choose a basis
$\{\ket i.\}_i$ in $\CC^d$ and let $\ket \Psi.:=\sum_i^d \ket
i.\otimes\ket i.$ be an (unnormalized) maximally entangled vector in
$\CC^d\otimes\CC^d$.  The object
\begin{equation}\label{eqn:choi}
	C_\Lambda:=(\Id\otimes \Lambda) \ket \Psi.\bra \Psi. 
	=\sum_{i,j} \ket i.\bra j. \otimes \Lambda(\ket i.\bra j.)
\end{equation}
is called the \emph{Choi matrix} of $\Lambda$. It is also known as the
\emph{process matrix} and the correspondence in Eq.\ (\ref{eqn:choi})
goes by the name of \emph{Jamio\l kowski isomorphism}. The name is
justified as $\Lambda \mapsto C_\Lambda$ is invertible:
\begin{equation}
	\Lambda(\ket i.\bra j.) = \bra i._1\,C_\Lambda\,\ket j._1.
\end{equation}
In what follows, we will write $T_{\mathcal{D}}$ for the channel induced by a
set of unitaries $\mathcal{D}$ via Eq.\ (\ref{eqn:uutwirling}) and denote the
corresponding Choi matrix by $C_\D$. Likewise, $C_{UU}$ designates the
Choi matrix of $T_{UU}$.

\subsection{The frame potential}
\label{sec:framePotential}

The various $\forall$-quantifiers in Definition \ref{def:design} make
it hard to identify a given set of matrices as a design. Any
exploration of this structure would thus greatly benefit from a simple
criterion for the property of  ``being a design''. Indeed, for the case
of spherical designs such a tool is well-known (see Ref.\
\cite{koenig} and references therein): a
set of vectors $\{\ket\psi_1.,\dots,\ket\psi_K.\}$ is a spherical
$2$-design in $\CC^d$ if and only if 
\begin{equation}\label{eqn:sphericalPotential}
	 \sum_{k,k'} |\braket \psi_k,\psi_{k'}.|^4 /K^2=2/(d^4+d^2).
\end{equation}
The expression on the left-hand side has been linked in Ref.
\cite{sics} to a concept which appeared in the context of
\emph{frame theory} in an equally insightful and enjoyable paper by
Benedetto and Fickus \cite{benedetto}. The authors considered a
physical model to introduce a notion of ``evenly distributed'' vectors:
if we assume that $K$ particles on the unit-sphere with respective
coordinates $\ket\psi_k.$ are subject to a repulsive force
proportional to $\braket \psi_k, \psi_{k'}.^2$, then the
left-hand-side of Eq.\ (\ref{eqn:sphericalPotential}) gives the
\emph{potential} of the configuration. Consequently, the quantity is
referred to as the (spherical) \emph{frame potential}
\footnote{
	More precisely, Eq.\ (\ref{eqn:sphericalPotential}) 
	is the
	\emph{second} frame potential \cite{sics}, the $t$-th one being
	induced by a repulsive force proportional to $\braket
	\psi_k,\psi_{k'}.^t$. As we will be concerned only with the second
	potential, we will drop the attribute from now on.
}. 
It turns out that $2/(d^4+d^2)$ is the lowest value the frame
potential can possibly attain and so there is a one-one correspondence
between global minimizers of the frame energy and spherical 2-designs.

Our first result transfers this nice concept to the setting of unitary
designs. 

\begin{theorem}[Frame potential]\label{thm:framePotential}
	Let $\D=\{U_k\}_{k=1,\dots, K}$ be a set of unitaries. Define the
	\emph{frame potential} of $\mathcal{D}$ to be
	\begin{equation}\label{eqn:framePotential}
		\mathcal{P}(\mathcal{D})= \sum_{U_k,U_{k'}\in\D} |\tr U_k^\dagger
		U_{k'}|^4 /K^2.
	\end{equation}
	The set $\mathcal{D}$ is a unitary 2-design if and only if
	$\mathcal{P}(\mathcal{D})=2$, which is a lower bound to the global
	minimum of the potential.
\end{theorem}

Theorem \ref{thm:framePotential} allows us to discuss the
connection between unitary and spherical designs quite explicitly.
Recall that the \emph{Hilbert-Schmidt inner product} on
$\mathcal{B}(\CC^d)$ is defined as $\braket A, B._{HS} := 
\tr(A^\dagger B)/d$. 
In the spirit of the Jamio\l kowski map, we can
establish an isomorphism between
$\mathcal{B}(\CC^d)$ as a vector space and $\CC^{d}\otimes\CC^d$.
Explicitly, we map
$U$ to $\ket v_U.$ where
\begin{equation}\label{eqn:vec}
	v_U^{ij} = \,{U^i}_j\, /\sqrt{d}.
\end{equation}
Here, we have used the notation $v^{ij} = (\bra i.\otimes\bra
j.)\,\ket v.$ and ${U^i}_j=\bra i. U \ket j.$ for the respective
matrix elements \footnote{
	The connection between Eq.\ (\ref{eqn:vec}) and the Jamio\l kowski
	isomorphism Eq. (\ref{eqn:choi}) is given by $\ket
	v_U.\bra v_U. = d^{-2}\, C_{U\spacedot U^\dagger}$.
}.
One checks that $\ket v_U.$ is a
normalized maximally entangled vector if and only if $U$ is unitary.
Hence, we can re-phrase Theorem \ref{thm:framePotential} as: $\D$ is a
unitary 2-design if and only if
\begin{equation}\label{eqn:vecPot}
	 \sum_{U_k,U_{k'}\in\D} |\braket v_{U_k}, v_{U_{k'}}.|^4/K^2 =2/d^4,
\end{equation}
which is the global minimum of the spherical frame potential for
$K$ \emph{maximally entangled} vectors. 
Note the close similarity to Eq. (\ref{eqn:sphericalPotential}).

The relation between unitary designs in $U(d)$ and
spherical designs in $\CC^d\otimes\CC^d$ now becomes apparent: both
correspond to minima of the frame potential, yet under different
constraints. For spherical designs the minimum is taken in the set of
all normalized vectors; whereas in the unitary case one demands that
the vectors are also maximally entangled.

Theorem \ref{thm:framePotential} also facilitates numerical searches
for designs on low dimensional spaces. Indeed, the authors have
written a program for the MatLab computer system, which numerically
minimizes the frame potential of a set of operators on $\CC^2$. If the
set has $K\geq 12$ elements, a multitude of unitary 2-designs is
found, while there seem to be no solutions for $K<12$. These findings
support Conjecture \ref{conj:cliffordBound}.

\begin{proof}\emph{(of Theorem \ref{thm:framePotential})}
	Using the notation introduced in Section
	\ref{sec:generalPreliminaries}, let $\Delta:=C_\D - C_{UU}$.
	Obviously, $\D$ is a 2-design if and only if $||\Delta||^2_2:=\tr
	|\Delta|^2=0$. We compute
	\begin{eqnarray*}	
		\tr(\Delta^\dagger \Delta) =\tr(C_{UU}^\dagger C_{UU} -
		C_{UU}^\dagger C_\D - C_\D^\dagger C_{UU} + C_\D^\dagger C_\D)
	\end{eqnarray*}
	and treat the terms in turn. To that end introduce a basis $\{\ket
	i.\}$ in $\CC^d\otimes \CC^d$ such that the first $d_s:=\dim P_S = d(d+1)/2$
	vectors are symmetric and the last $d_a:=\dim P_A=d(d-1)/2$ ones
	anti-symmetric with respect to $\FF$. In the formulas below, we will
	sometimes write $\ket i_S.$ or $\ket i_A.$ to indicate the subset a
	given vector belongs to.  Note that the vector $\ket\Psi.$
	introduced in Section \ref{sec:generalPreliminaries} can be written
	as $\ket\Psi.=\sum_{i_s}\ket i_S.\otimes\ket i_S.+\sum_{i_A} \ket
	i_A.\otimes \ket i_A.$. One then finds
	\begin{eqnarray*}
		C_{UU}
		&=& 
		\sum_{i_S,j_S} \ket i_S. \bra j_S. \otimes
		\tr(\ket i_S. \bra j_S. P_S) P_S'+
		S \leftrightarrow A \\
		&=&P_A \otimes P_A' + P_S \otimes P_S'.
	\end{eqnarray*}
	We have used the abbreviation $S\leftrightarrow A$ to denote the
	term which follows from the preceding one by a straight-forward
	substitution of symmetric by anti-symmetric expressions, and as before
	$P_S' = P_S/\tr P_S$, $P_A' = P_A/\tr P_A$.
	Further:
	\begin{eqnarray*}
		C_\D &=&
		\sum_{i,j} \ket i. \bra j. \otimes 
		\left( \sum_k U_k^{\otimes 2} \ket i. \bra j.
		(U_k^\dagger)^{\otimes 2} /K\right).
	\end{eqnarray*}
	Hence,
	\begin{eqnarray*}
		\tr(C_{UU}^\dagger C_{UU})
		&=&d_S^{-2} \tr (P_S\otimes P_S) + d_A^{-2} \tr (P_A\otimes P_A)=2
	\end{eqnarray*}
	and
	\begin{eqnarray*}
		&&\tr(C_{UU}^\dagger C_\D) \\
		&=& K^{-1}
		\sum_{i,j} \tr(P_S \ket i. \bra j.) \, 
		\tr \left(P_S' \sum_k U_k^{\otimes 2} \ket i. \bra j.
		(U_k^\dagger)^{\otimes 2}\right) \\
		&&+S\leftrightarrow A\\
		&=& K^{-1}
		\sum_{i_S} 
		\tr \left( \sum_k U_k^{\otimes 2} P_S' \ket i_S. \bra i_S.
		(U_k^\dagger)^{\otimes 2}\right) 
		+S\leftrightarrow A\\
		&=& 
		\sum_{i_S} \tr (P_S' \ket i_S. \bra i_S.) +S\leftrightarrow A\\
		&=& 2=\tr(C_\D^\dagger C_{UU}).
	\end{eqnarray*}
	Lastly:
	\begin{eqnarray*}
		&&\tr(C_\D^\dagger C_\D)\\
		&=&
		K^{-2} \sum_{i,j} \sum_{k,k'} \tr\left(
		U^{\otimes 2}_k \ket j. \bra i. (U^\dagger_k)^{\otimes 2}\,\,
		U^{\otimes 2}_{k'} \ket i. \bra j. (U^\dagger_{k'})^{\otimes 2}
		\right)\\
		&=&
		K^{-2} \sum_{k,k'} |\tr\,U_k^\dagger U_{k'}|^4=\mathcal{P}(D).
	\end{eqnarray*}
	The claim is now immediate.
\end{proof}

For the construction of approximate unitary designs in Section
\ref{sec:approximations}, we record the following corollary.

\begin{corollary}\label{cor:choiDistance}
	Let $\D$ be a set of unitary matrices, $C_{UU}$ and $C_\D$ as
	defined in Section \ref{sec:generalPreliminaries}.  Then
	\begin{equation*}
		||C_{UU}-C_\D||_2^2 = \mathcal{P}(\mathcal{D})-2.
	\end{equation*}
\end{corollary}

\subsection{A lower bound}
\label{sec:bounds}

Intuitively it is clear that constructing unitary designs becomes more
challenging the fewer elements $K$ one allows for (c.f. Theorem
\ref{thm:groupDesigns}.\ref{itm:supergroups}).  This section is
devoted to finding a lower bound for $K$ as a function of the
dimension $d$.

What is the situation to date? Before the present paper, all known
families of 2-designs were subgroups of the Clifford group
$\mathcal{C}_d$ in prime-power dimensions $d$. In the context of
quantum information, the Clifford group is the set of unitaries
mapping the set of Weyl operators (also known as: generalized Pauli
operators) to itself under conjugation \cite{gottesman}. An
introduction into this theory will be given in Section
\ref{sec:cliffordDesigns}, where all claims made in this paragraph
will be elaborated on. References
\cite{dataHiding,duer2,dankert,dankertThesis} use the fact that the
full Clifford group $\mathcal{C}_d$ constitutes a 2-design. However,
as the cardinality of $\mathcal{C}_d$ grows exponentially in $d$ (c.f.
Eq.\  (\ref{eqn:cliffordCardinalities})), one might hope for the
existence of more optimal designs. Fortunately, in Ref.\ \cite{chau} it
has been realized that a particular subgroup of the Clifford group already
possesses the 2-design property. The group's order scales as $O(d^5)$.
What is more, the existence of Clifford 2-designs with
$d^2(d^2-1)=O(d^4)$ elements has been established for several
dimensions \cite{chau}. As will be explained in Section
\ref{sec:cliffordDesigns}, $d^2(d^2-1)$ is in fact the smallest value
a design based on the Clifford group can possibly have and that value
will subsequently be referred to as the \emph{Clifford bound}. For
various reasons to be explained later, we believe this to be a general
lower bound for the cardinality of any 2-design, even for
constructions which are not based on the Clifford group.

\begin{conjecture}\label{conj:cliffordBound}
	The Clifford bound 
	\begin{equation}
		d^4-d^2
	\end{equation}
	is a lower bound for the cardinality of any unitary 2-design.
\end{conjecture}

While we were not able to prove this conjecture, an estimate which
equals the Clifford bound in leading order is established below.

\begin{theorem}[Lower bound on $K$]\label{thm:lowerBound}
	A unitary 2-design in dimension $d$ has no fewer than 
	\begin{equation}
		d^4-2d^2+2
	\end{equation}
	elements.
\end{theorem}

Note that a spherical 2-design in $\CC^{d}\otimes\CC^d$ has at least
$d^4$ elements -- so the slightly higher frame potential
characteristic for unitary 2-designs might allow one to save a few
elements as compared to spherical 2-designs.

\begin{proof}
	We follow an idea from Ref.\ \cite{sics}. Let $\ket
	v_U.:= U\otimes\Id\,\ket v_0.$ for some maximally entangled vector
	$\ket v_0.$. Define a homogeneous polynomial $p$ of degree $2,2$
	in $U$ by
	\begin{equation}\label{eqn:lowerBoundPolynom}
		p(U):=  \bra v_U. A \ket v_U. \tr (\ket v_U.\bra v_U.\,B).
	\end{equation}
	Certainly, Eq.\ (\ref{eqn:design}) holds and hence, as $A$ and $B$
	are arbitrary, the relation
	\begin{eqnarray}\label{eqn:gChannel}
		&& \sum_{U\in{\cal D}} \bra v_U. A \ket v_U. \, \ket v_U. \bra v_U. /K\\
		&=&	
		\int \bra v_U. A \ket v_U. \, \ket v_U. \bra v_U.\,dU
		=:\Lambda(A) \nonumber
	\end{eqnarray}
	must hold and defines a channel $A\mapsto \Lambda(A)$. We want to
	compute the kernel of $\Lambda$. 

	The channel $\Lambda$ is clearly $U\otimes \Id$-covariant:
	\begin{equation}
		\Lambda((U\otimes \Id) A (U \otimes \Id)^\dagger )
		= (U\otimes\Id) \, \Lambda(A) \, (U\otimes\Id )^\dagger. 
	\end{equation}
	But because for any maximally entangled state $(\Id\otimes V) \ket
	v.=(V'\otimes\Id) \ket v.$ for some $V'$, 
	$\Lambda$ is also $\Id\otimes V$ and
	hence even $U\otimes V$-covariant, for all unitaries $U,V$. Invoking
	Schur's Lemma one concludes that $\Lambda$ must be a mixture of
	projections onto the $U\otimes V$-invariant subspaces of
	$\mathcal{B}(\H)$. What are these spaces? Let us first identify the
	irreducible components of $U\spacedot U^\dagger$. The multiples of the
	identity ($M_1$ for short) clearly form an irreducible component by
	themselves. Its complement is the space of trace-less operators
	($M_2$). Now, $U\spacedot U^\dagger$ must act irreducibly on $M_2$,
	because there exists bases of mutually conjugate trace-less
	operators
	\footnote{
		Take e.g. $A_{i,j}=\ket i.\bra j.+\ket j.\bra i., B_{i,j}=i(\ket
		i.\bra j. - \ket j. \bra i.)$. For $i\neq j$, all operators 
		$A_{i,j},
		B_{i,j}$ have the same set of eigenvalues 
		$1,-1,0,0,\dots$ and are
		thus mutually conjugate. These operators clearly 
		form a basis in
		the space of trace-less observables.
	}.
	
	Surely then, $M_1\otimes M_1$ (multiples of the
	identity), $M_1\otimes M_2, M_2\otimes M_1$ (the local observables
	of the form $\Id\otimes X, X\otimes \Id$) and $M_2\otimes M_2$ are
	invariant under $U\otimes V\spacedot U^\dagger\otimes V^\dagger$. A
	moment of thought reveals that they are irreducible (think of cyclic
	vectors). Clearly, $M_1\otimes M_1$ has no non-trivial intersection
	with $\ker(\Lambda)$, while
	$M_1\otimes M_2, M_2\otimes M_1 \subset \ker(\Lambda)$. What about
	$M_2\otimes M_2$? Because of $\Lambda$'s structure, either any
	element of $M_2\otimes M_2$ is in the kernel or else, none is. 
	Setting $B=A^\dagger$ and evaluating Eq.\ (\ref{eqn:lowerBoundPolynom})
	we see that 
	\begin{equation}
		p(U)=|\bra v_U. A \ket v_U.|^2\geq 0, 
	\end{equation}
	so
	$\Lambda(A)=0$ if and only if $\bra v. A \ket v.  =0$ for all
	maximally entangled vectors $\ket v.$. To conclude that $M_2\otimes
	M_2$ has no intersection with $\ker \Lambda$, we only need to
	assure the existence of a single traceless observable $X$ and a single
	maximally entangled state $\ket v.$ such that $\bra v. X\otimes X\ket
	v.\neq 0$, which is trivially possible.
	Hence 
	$\rank \Lambda=d^4-\dim\ker\Lambda=d^4 - \dim(M_1\otimes
	M_2)-\dim(M_2\otimes M_1)=d^4-2(d^2-1)$. 
	But the rank of $\Lambda$
	cannot be larger than $K$, by Eq.\ (\ref{eqn:gChannel}).
\end{proof}

\section{Group designs}
\label{sec:groupDesigns}

When searching for unitary designs, it might prove helpful to assume some
additional structure in order to narrow down the search space and
simplify the proofs. Indeed, sets of unitary matrices appear most
naturally as representations of finite groups and (except for our
numerical findings), all known designs are matrix
groups. It will turn out that the concept of unitary designs has a
very natural interpretation in terms of representation theory.

\subsection{Irreducible constituents}
\label{sec:irreducibleConstituents}

We will be concerned with sets $\D$ of unitaries which form a finite
matrix group on $\CC^d$. It will prove convenient to conceive $\D$ as
the image of a representation $U: g\mapsto U_g$ of some finite group
$G$.  Recalling the notions of Section \ref{sec:generalPreliminaries},
it is clear that the channel $T_\D$
is nothing but the twirling channel associated with the representation
$U$.  By Eq.\  (\ref{eqn:generaltwirling}), $T_\D$ will project onto
the irreducible subspaces of this representation.  As any operator of
the form $U_g\otimes U_g$ commutes with the flip operator $\FF$, we
know that the symmetric and anti-symmetric subspaces of $\CC^d\otimes
\CC^d$ will be among the invariant subspaces of $\{U_g\otimes
U_g\,|\,g\in G\}$. In general, these spaces are not
going to be irreducible. We now see what makes representations $U$
which induce a 2-design special: \emph{ $T_\D=T_{UU}$ (and hence $\D$
is a design) if and only if the representation $g\mapsto U_g\otimes
U_g$ has exactly two irreducible components.}

Simple as this observation may be, it must not be underestimated: it
allows us to understand designs from a group theoretical point of
view. The next section will further elaborate on this approach.

\subsection{Characters}
\label{sec:characters}

Let us devote one paragraph to recall some very basic notions and
results from representation theory \cite{isaacs}. To every unitary
representation $U: g\mapsto U_g$ of a finite group, one associates its
\emph{character} $\zeta(g)=\tr U_g$. One says that the representation
\emph{affords} $\zeta$. Denote the irreducible representations
(\emph{irreps}) of $G$ by $\{V^{(i)}\}_i$ and their associated
\emph{irreducible characters} by $\{\chi_i\}$.  One introduces a
scalar product between characters by setting
\begin{equation}
	\langle \zeta, \chi \rangle := |G|^{-1} \sum_g \bar\zeta(g) \chi(g).
\end{equation}
It is a well-known and fundamental relation that the irreducible
characters are ortho-normal: $\langle \chi_i, \chi_j
\rangle=\delta_{i,j}$. The fact that any representation reduces to a
direct sum of irreps means that any character can be expanded in terms
of the irreducible ones and further that $\langle \zeta, \chi_i
\rangle$ gives the number of times $n_i$ the $i$-th irrep occurs in
the decomposition of the representation affording $\zeta$. Finally, if
$\zeta=\sum_i n_i \chi_i$, then $||\zeta||^2=\langle \zeta, \zeta
\rangle=\sum_i n_i^2$. 

Now, let $\D,G,U$ be as in Section \ref{sec:irreducibleConstituents}.
We compute the frame potential of $\D$:
\begin{eqnarray}
	\mathcal{P}(\D) 
	&=&  \sum_{g,g'} |\tr U_{g^{-1} g'}|^4 /|G|^2\\
	&=& \sum_{g} |\tr U_g |^4  /|G| \nonumber \\
	&=&   \sum_{g} \overline{(\tr U_g\otimes U_g)} (\tr U_g\otimes U_g) /|G| \nonumber \\
	&=& \langle \zeta_{U^{(2)}}, \zeta_{U^{(2)}} \rangle 
	= ||\zeta_{U^{(2)}}||^2, \nonumber
\end{eqnarray}
where $\zeta_{U^{(2)}}(g)=\tr (U_g\otimes U_g)=\zeta_U(g)^2$ is the
character of the representation $U^{(2)}: g\mapsto U_g\otimes U_g$. In
other words, the frame potential of a group design is the squared norm
of the character of $U^{(2)}$.

We can now rederive Theorem \ref{thm:framePotential}. By this
section's first paragraph, $||\zeta_{U^{(2)}}||^2=\sum_i n_i^2$, which
equals $2$ if and only if $U^{(2)}$ has exactly two irreducible
components. This in turn is equivalent to $U$ inducing a 2-design, as
has been shown in Section \ref{sec:irreducibleConstituents}. Note how
much the group structure simplified the proof.

It seems remarkable that the frame potential offers a very natural
interpretation in terms of two completely unrelated structures: from
the point of view of frame theory, it is a purely \emph{geometrically}
motivated measure for the ``eveness'' of a distribution. In terms of
group representation theory, it seemlessly takes on an
\emph{algebraic} role.

\subsection{General results and properties}
\label{sec:groupGeneral}

Consider a group design $\D=\{U_g\,|\,g\in G\}$. The \emph{center}
$\mathcal{Z}(U)$ of $U$ are the elements of $\D$ which commute with
any $U_h$. By Schur's Lemma, if $U$ is irreducible, we have that
$U_g\in \mathcal{Z}(U)\Leftrightarrow U_g \propto \Id$. Now choose one
representative of each coset $\D/\mathcal{Z}(U)$ and assemble these
unitaries in a set $\D'$ ($\D'$ is called a \emph{transversal} of
$\D/{\mathcal{Z}(U)}$). Using Eq.\ (\ref{eqn:uutwirling}), one
sees that $D'$ is a 2-design of cardinality $|\D|/|\mathcal{Z}(U)|$.
From now on, we will restrict attention to such reduced sets.
Consequently, for any representation $U$ of $G$, we will define $\D_U$
to be a transversal of $\{U_g\,|\,g\in G\}/\mathcal{Z}(U)$ and refer
to $\D_U$ as the \emph{group design induced by $U$}.

Using this definition, let us collect and extend the results on group
designs in the following theorem.

\begin{theorem}[Group designs]
	\label{thm:groupDesigns}
	Let $G$ be a finite group and $U$ a unitary representation of $G$ on
	$\CC^d$ affording the character $\zeta$. The following are
	equivalent:
	\begin{enumerate}
		\item
		The set $\D_U$ is a 2-design.

		\item
		The representation $U^{(2)}: g\mapsto U_g\otimes U_g$ has no more
		than two irreducible components.

		\item
		It holds that $||\zeta_{U^{(2)}}||^2=2$.

		\item
		The characters
		\begin{eqnarray*}
			\chi_S(g)&:=&(\chi(g)^2+\chi(g^2))/2,\\
			\chi_A(g)&:=&(\chi(g)^2-\chi(g^2))/2\\
		\end{eqnarray*}
		are irreducible.
	\end{enumerate}
	Further:
	\begin{enumerate}
		\setcounter{enumi}{4}
		\item\label{itm:cardinality}
		The cardinality $K=|\D_U|$ is a multiple of $d$ and $1/2d
		(d\pm1)$.

		\item\label{itm:supergroups}
		Let $H$ be a finite group represented on $\CC^d$ by V. If
		$\{V_h\,|\,h\in H\}\supset \{U_g\,|\,g\in G\}$ then
		$\mathcal{F}(\D_V)\leq\mathcal{F}(\D_U)$.

		\item\label{itm:quantizedPotential}
		The frame potential of a matrix group is an integer. 

		\item
		A necessary condition for $\D_U$ to be a 2-design is that $U$ is
		irreducible.

		\item
		For $d>2$, there are no 
		real-valued group representations which
		form a
		2-design.
	\end{enumerate}
\end{theorem}

Statement \ref{itm:cardinality} can be used in conjunction with
Section \ref{sec:bounds} to derive bounds on $K$. For example, for
$d=2$, it holds that $K\geq 10$ by Theorem \ref{thm:lowerBound}. But
we now know that $K$ must be divisible by $2$ and $3$, so that $K\geq
12$. As unitary group designs of order $12$ in dimension $2$ do indeed
exist, we know that the bound is tight. Unfortunately, this is the
only case where we can make lower and upper bounds match. Note also,
that $12$ is the value predicted by the Clifford bound for $d=2$,
supporting our conjecture.

The \ref{itm:supergroups}-th point says that ``supergroups have lower
frame potential than their subgroups''. Again, it is clear that
constructing designs is easier, the more elements one allows for. In
general, however, just adding further unitaries to an ``almost design''
is not going to improve the potential. For group-designs the situation
is different, as we now know.

Lastly, statement \ref{itm:quantizedPotential} says that the frame
potential of matrix group is ``quantized''. 
In that sense, \emph{there are no  ``approximate group designs''}.

\begin{proof}
	The equivalence {\it 1.} $\Leftrightarrow$ {\it 2.}
	$\Leftrightarrow$ {\it 3.} has been established in the discussion
	preceding the theorem. Claim {\it 4.} is equivalent to {\it 2.}, as
	$\chi^2=\chi_S+\chi_A$. The fifth statement follows from a
	well-known theorem in representation theory (see Ref.\ 
	\cite{isaacs}). 
	Point {\it 6.} holds true as the number of
	irreducible components cannot decrease when passing from a subgroup
	to a supergroup. Claims {\it 7.} and {\it 8.} should be obvious. Lastly, {\it
	9.} is valid because for real $\zeta_U$
	\begin{eqnarray*}
		1=\langle \zeta_U,\zeta_U\rangle = 
		\langle \zeta_U \bar \zeta_U, 1_G\rangle = \langle \zeta_{U^{(2)}},
		1_G\rangle,
	\end{eqnarray*}
	where $1_G: g \mapsto 1$ is the trivial representation. Hence, $1_G$
	is a one-dimensional irreducible component of $U^{(2)}$. But for
	$d>2$ we have that $d_s, d_a \neq 1$.
\end{proof}

\subsection{Harvesting character tables}
\label{sec:characterTables}

The results of Section \ref{sec:characters} enable us to identify
designs by just looking at character tables of finite groups. Such
tables have been the subject to intensive research and are digitally
available. We have employed the freely available GAP computer system
\cite{gap} to search the GAP Character Table Library version 1.1
\cite{ctllib} for unitary designs. Some findings are compiled in Table
\ref{tbl:gap}. For each dimension $d$ in which a unitary design has
been found, one example is included in the table. To access the listed
character tables, pass the name to \verb CharacterTableFromLibrary() .
The column ``Irred.\ character no.'' gives the position of the design
within the list of irreducible characters returned by the \verb.Irr().
-function. E.g., the dialog
\begin{verbatim}
gap> t:=CharacterTableFromLibrary("J4");;
gap> chr:=Irr(t)[2];;                    
gap> Degree(chr);
1333
gap> Norm(chr*chr);
2
gap> 
\end{verbatim}
confirms that the last item in Table \ref{tbl:gap} does indeed belong
to a unitary group design in dimension $1333$.

\section{Phase space techniques}
\label{sec:phaseSpace}

The title \emph{phase space techniques} refers to any method employing
the realted concepts of 
Weyl operators (also known as \emph{generalized Pauli operators}),
stabilizer states and the Clifford group. These structures have played
a central role in the theory of both spherical and group designs
\cite{sics,mubs,dataHiding,dankert,chau}. As
the following paragraphs require some rather technical preparations,
we state a summary of the results at this point.

In Section \ref{sec:approximations} \emph{asymptotic} unitary designs
will be constructed. By this, we understand a family of sets of
unitaries $\D_d$, such that the matrices in $\D_d$ are $d$-dimensional
and $\lim_{d\to\infty} \mathcal{P}(\D_d) =2$.  The intuition behind
the construction is as follows: in Section \ref{sec:framePotential},
we discussed the relation between the frame potential of operators on
$\CC^d$ and vectors in $\CC^{d^2}$. Hence it is natural to ask whether
one can exploit this relation to turn spherical designs into unitary
ones.  Obviously, in order to obtain \emph{unitary} matrices, we must
require the vectors in the spherical design to be maximally entangled.
Recall that a maximal set of MUBs is a 2-design and, moreover, that
such sets can be chosen to consist of stabilizer states \cite{mubs}.
For bi-partite systems, where each party has prime dimension, it is
known that stabilizer states are either maximally entangled or not
entangled at all \cite{stabilizerEnt}.  It is thus reasonable to
assume that among the elements of a maximal set of MUBs, there are
``enough'' maximally entangled ones to yield a set of unitaries with a
low frame potential.  Fortunately, this intuition turns out to be true
and we will find sets of $O(d^4)$ unitaries in dimension $d=p^n$,
which approximate a 2-design as $d\to\infty$.

Secondly, in Section \ref{sec:cliffordDesigns}, we will revisit the
technique of Clifford twirling. Our main contribution to the theory
will be a systematical reassessment of what is already known. Indeed,
reading the literature, one gets the impression that some confusion
has arisen due to the fact that \emph{several distinct Clifford groups
exist}. The one used in Refs.\
\cite{dankert,dankertThesis,dataHiding,duer2} is different from the
one in Ref.\ \cite{chau}. Going on, we will review a construction by
Chau, which meets the Clifford bound in dimensions $2,3,5,7,11,$ and
outline a way for circumventing a no-go theorem which asserts that for
any other dimension the bound cannot be met. In particular, for $d=9$,
we present a subgroup of the Clifford group which is a 2-design of
smaller cardinality than Ref.\ \cite{chau} seems to suggest is
possible. 

Before presenting these results in detail, the reader must endure the
tour-de-force of technical preparations given in Section
\ref{sec:phaseSpaceIntro}. It is a peculiarity of the theory to be
presented that it works much more smoothly in odd dimensions $d$ than
in even ones. While it can be checked that all results in the next
section also hold for the qubit case, the proofs are given only for
the case of odd $d$.

\subsection{Introduction}
\label{sec:phaseSpaceIntro}

This section contains a very brief outline of the general theory. See
Ref.\ \cite{poswig,diploma} and references therein for a more detailed
exposition. 

\subsubsection{Weyl operators, the Jacobi group \& the Clifford group}
\label{sec:weylAndCo}

Let us first gather some well-known facts on finite fields
\cite{jungnickel}. If $p$ is prime and $r$ a positive integer,
$\FF_{p^m}$ denotes the unique finite field of order $p^m$. The
simplest case occurs for $m=1$, when $\FF_p\simeq \ZZ_p$, i.e., the
set of integers \emph{modulo} $p$. 
Now set $d=p^m$ and choose an $r\in\NN$.
Out of the \emph{base field}
$B:=\FF_d$, one can obtain the fields $\FF_{d^r}$ by means of a
\emph{field extension}.
Extension fields contain the base field as a
subset.  
The extension field possesses the structure of an
$r$-dimensional vector space over the base field. A set of elements
of $F$ is a \emph{basis} if it spans the entire field under addition
and $B$-multiplication. 
The operation
\begin{equation*}
	\Tr_{F/B} f = \sum_{k=0}^{r-1} f^{d^k}
\end{equation*}
takes on values in the base field and is $B$-linear. Therefore, 
\begin{equation*}
	\tuple f,g. \mapsto \Tr_{F/B} (fg)
\end{equation*}
defines a $B$-bilinear form. For any basis $\{b_i\}$,
there exists a \emph{dual basis} $\{b^i\}$ fulfilling the
relation
	$\Tr_{F/B} (b^i b_j) = \delta_{i,j}$
(we do not use Einstein's summation convention). Clearly, if $f\in F$
can be expanded as $f=\sum_i f^i b_i$, with coefficients $f^i \in B$,
then duality implies that $f^i = \Tr (f b_i)$.

We will work in the $d:=p^m$-dimensional Hilbert space $\mathcal{H}\simeq
\CC^d$ spanned by the vectors $\{\ket a. \, | \, a\in \FF_d\}$.
Define a \emph{character} of $\FF_d$ by $\chi_d(a):=\exp(i \frac{2\pi}p
\Tr_{\FF_{p^m}/{\FF_p}}(a))$.
The relations
\begin{eqnarray}\label{shiftClock}
  \Hat x_d(q)\ket x. = \ket x+q., \quad\quad
	\Hat z_d(p)\ket x. = \chi_d(p x) \ket x.
\end{eqnarray}
define the \emph{shift} and \emph{boost} operators respectively.  The
Weyl operators (also known as \emph{generalized Pauli operators}) in
dimension $d$ are given by
\begin{eqnarray}\label{eqn:weylOps}
	w_d(p,q)=\chi_d(-2^{-1}p q)\, \Hat z_d(p)\Hat x_d(q),
\end{eqnarray}
for $p,q\in \FF_d$. The phase factors in Eq.\
(\ref{eqn:weylOps}) have been included to clean up some later
formulas. The \emph{phase space} $V$ is defined as
$V:=\FF_{d} \times \FF_{d}$. We introduce the standard
\emph{symplectic inner product} on $V$ by
\begin{eqnarray}\label{eqn:symp}
  \Symp p,q,p',q'.:=pq'-qp'.
\end{eqnarray}
For elements $a=(p,q)^T$ of $V$, we set $w_d(a):=w_d(p,q)$. Denote by
$\mathcal{W}_d := \{w_d(a)\,|\,a\in V\}$ the
collection of all Weyl operators. 
The \emph{commutation relations}
\begin{equation}\label{eqn:commutation}
	w_d(a) w_d(b) = \chi_d([a,b]) w_d(b) w_d(a),
\end{equation}
can be checked to hold.

Let $S$ be a symplectic $2\times 2$ matrix
with entries in $\FF_d$. There exists a unitary operator $\mu_d(S)$
defined via
\begin{equation}\label{eqn:meta}
	\mu_d(S)\,w_d(a)\,\mu_d(S)^\dagger = w_d(S a)
\end{equation}
for all $a\in V$. We will call $\mu_d(S)$ the \emph{metaplectic
representation} of $S$. 
Up to phase factors, the set of unitaries of the form $\mu_d(S)\,w_d(a)$
constitute a group, which will be referred to as the \emph{Jacobi group}
$\mathcal{J}_{d}$.

The preceding definition have been made with a single $d$-dimensional
particle in mind. We now consider the situation of $n$ particles, each
having $d=p^m$ levels. The Hilbert space becomes $\CC^{d^n}$ spanned
by $\{\ket a. \,|\, a \in \FF_d^n \}$. Let $p=(p_1,\dots,p_n),
q=(q_1,\dots,q_n)$. We define the Weyl operators as
\begin{equation}\label{eqn:multiWeyl}
	w_{d,n}(p,q)=w_d(p_1,q_1)\otimes\dots\otimes w_d(p_n,q_n).
\end{equation}
In this case, the phase space is set to be $V=\FF_d^n\times \FF_d^n$
and Eqs. (\ref{eqn:symp},\ref{eqn:commutation}) continue to make sense
if we perceive products between elements of $p,q\in \FF_d^n$ as a
canonical scalar product: $pq=\sum_i^n p_i q_i$. In complete analogy
to the $n=1$ case, one finds that for any symplectic $2n\times
2n$ matrix $S$ with entries in $\FF_d$, there exists an operator
$\mu_{d,n}(S)$ such that
\begin{equation}\label{eqn:multiMeta}
	\mu_{d,n}(S)\,w_{d,n}(a)\,\mu_{d,n}(S)^\dagger = w_{d,n}(S a)
\end{equation}
holds for all $a\in V$. Denote the set of Weyl operators according to
Eq.\ (\ref{eqn:multiWeyl}) by $\mathcal{W}_{d,n}$ and the Jacobi group
spanned by $\{w_{d,n}(a) \mu_{d,n}(S)\}_{a,S}$ by $\mathcal{J}_{d,n}$.

For a Hilbert space of prime-power dimension $p^s$, we can now
construct an entire family of different Weyl operators and Jacobi
groups. Indeed, for any $n,m$ such that $nm=s$, the Weyl operators
$w_{p^m,n}$ are $p^s$ dimensional. 
Prominent choices include $n=1,m=s$
(used in Ref.\ \cite{chau}) and $n=s,m=1$ (used in Refs.
\cite{dankert,dataHiding,duer2}). It will turn out that all
definitions of the \emph{Weyl operators} coincide, while the various
\emph{Jacobi groups} differ.  Proposition \ref{prop:relations} makes
these remarks precise.  In order to state it, we need one final
definition: the \emph{Clifford group} $\mathcal{C}_{p,n}$ is the set
of unitaries mapping the set $\mathcal{W}_{p,n}$ onto itself under
conjugation. This definition reflects the general use of word
\emph{Clifford group} in quantum information theory \cite{clifford}.

\begin{proposition}\label{prop:relations}
	Let $p^s$ be a power of a prime. Let $n<n'$ and $m,m'$ be such
	that $mn=m'n'=s$.
	Then
	\begin{eqnarray}
		\mathcal{W}_{p^m,n} &=& \mathcal{W}_{p^{m'},n'}, \label{eqn:weyls}\\ 
		\mathcal{J}_{p^m,n} &\subset& \mathcal{J}_{p^{m'},n'}. 
		\label{eqn:jacobiRelations} 
	\end{eqnarray}
	The inclusion in Eq.\ (\ref{eqn:jacobiRelations}) is
	proper and $\mathcal{J}_{p,s}=\mathcal{C}_{p,s}$.
\end{proposition}

For the construction in Section \ref{sec:approximations}, it will be
necessary to understand Eq.\ (\ref{eqn:weyls}) in more detail. Indeed,
it has been realized before \cite{diploma,pittenger,poswig} that the Weyl operators in
$\mathcal{W}_{p^n,1}$ can be written as tensor products of those in
$\mathcal{W}_{p,n}$. In what follows, we will refine this picture.

Let $d=p^m$ be a power of a prime, let $B=\FF_d$. Let $F=\FF_{d^n}$
be an extension field of $B$. In $F$, choose a basis
$\{b_i\}_{i=1\dots n}$ over $B$. Denote the dual basis by
$\{b^i\}_i$. Having general relativity conventions in mind, we will adopt the
following notation: for an element $f\in F$, we denote its expansion
coefficients with respect to $b_i$ by $f^i$ and the coefficients for
the dual bases by $f_i$:
\begin{equation}
	f = \sum_i f^i b_i = \sum_i f_i b^i.
\end{equation}
The Weyl operators in $\mathcal{W}_{d,1}$ act on $\H=\CC^{d^n}\simeq
(\CC^d)^{\otimes n}$, where we choose the isomorphism to be
implemented by
\begin{equation}
	\ket q. = \ket q^1 b_1+ \dots +q^n b_n. \mapsto \ket
	q^1.\otimes\dots\otimes\ket q^n..
\end{equation}

\begin{lemma}[Factoring Weyl operators]
	Using the notions introduced above, the Weyl operators in
	$\mathcal{W}_{d^n,1}$ factor as
	\begin{equation}
		w_{d^n}(p,q)=w_{d}(p_1,q^1)\otimes\dots\otimes
		w_{d}(p_n,q^n).
	\end{equation}
\end{lemma}

\begin{proof}
	Denote the common \emph{prime field} $\FF_p$ of $B$ and $F$ as $P$.
	It is well-known \cite{jungnickel} that
	$\Tr_{B/P}\circ\Tr_{F/B}=\Tr_{F/P}$.  Hence 
	\begin{eqnarray*}
		\chi_F(p q) 
		&=& \chi_B\big(\sum_{i,j} p_j q^i \Tr_{F/B}(b_i b^j)\big)
		= \prod_i \chi_B(p_i q^i).
	\end{eqnarray*}
	Similarly,
	\begin{eqnarray*}
		\Hat x_F\big( \sum_i q^i b_i \big) \ket \sum_j x^j b_j.
		&=& \ket \sum_i (q^i+x^i) b_i . \\
		&=& \bigotimes_i\Hat x_B(q^i) \ket x^i. ,\\
		\Hat z_F\big( \sum_i p_i b^i \big) \ket \sum_j x^j b_j. 
		&=& \prod_i \chi_B(p_i x^i) \ket \sum_j x^j b_j. \\
		&=& \bigotimes_i\Hat z_B(p_i) \ket x^i.. \nonumber
	\end{eqnarray*}
	Using Eq.\ (\ref{eqn:weylOps}), the claim follows.
\end{proof}

\begin{proof}\emph{(of Proposition \ref{prop:relations})}
	Eq.\ (\ref{eqn:weyls}) follows from the previous lemma.  Saying
	that $\mathcal{J}_{p,r}=\mathcal{C}_{p,r}$ is just rephrasing the
	definition of the Clifford group. For Eq.
	(\ref{eqn:jacobiRelations}) the reader is deferred to Refs.
	\cite{poswig,diploma}.
\end{proof}

\subsubsection{Stabilizer states}
\label{sec:stabilizers}

Using the commutation relations Eq.\ (\ref{eqn:commutation}) it is
immediate that to Weyl operators $w(a), w(b)$ commute if and only if
$[a, b]=0$. Now consider the image of an entire subspace $M$ of $V$
under $w$:
\begin{equation*}
	w(M)=\{w(m)|m\in M\}.
\end{equation*}
The latter set consists of commuting operators if for all
$m_1, m_2\in M$, the symplectic inner product vanishes:
$[m_1,m_2]=0$. If that condition is fulfilled, $w(M)$ is called a
\emph{stabilizer group}. Consider the operator
\begin{equation}
	\rho_M:= \sum_{m\in M} w(m)/|M|.
\end{equation}
One checks that
\begin{eqnarray}\label{eqn:projector}
	\tr \rho_M
	&=&   \sum_m \tr w(m) /|M| \\
	&=&    \sum_m d \delta_{m,0} /|M| = d/|M|.\nonumber
\end{eqnarray}
and, using the fact that $M$ is a linear space,
\begin{eqnarray}\label{eqn:idempotent}
	\rho_M\rho_M 
	&=&  |M|^{-2} \sum_{m,m'} w(m+m') \\
	&=& |M|^{-1} \sum_m w(m) = \rho_M.\nonumber
\end{eqnarray}
Hence, if $|M|=d$, then $\rho_M=\ket\psi_M.\bra\psi_M.$ is a rank-one projector 
and $\ket\psi_M.$ is called the \emph{stabilizer state} associated
with
$M$. The preceding definition can be extended: choose a character
$\zeta$ of $M$ (i.e., a function $M\to\CC$ such that
$\zeta(m_1+m_2)=\zeta(m_1)\zeta(m_2)$) and set 
\begin{equation}\label{eqn:generalStabilizer}
	\rho_{M,\zeta}:=\sum_m \zeta(m) w(m) /|M| .
\end{equation}
The calculations Eqns.\ (\ref{eqn:projector},\ref{eqn:idempotent}) can
be repeated and one finds that also $\rho_{M,\zeta}$ projects onto a
vector, which will be denoted by $\ket\psi_{M,\zeta}.$. 

\subsubsection{Mutually unbiased bases}
\label{sec:mubs}

We will recall a well-known construction for MUBs
\cite{mubs}. Once again, let
$F=\FF_{p^m}$ be a finite field and $V=F^2$ the associated phase
space. Let
\begin{equation}
	v_a = \Vec a,1.,\quad\quad M_a=\{\lambda v_a\,|\,a\in F\}.
\end{equation}
Clearly, $M_a$ is a one-dimensional subspace of $V$ and hence of
cardinality $|M_a|=|F|=d$. Because the symplectic form is
anti-symmetric $[a,b]=-[b,a]$ it holds for $\lambda v_a, \lambda' v_a
\in M_a$ that $[\lambda v_a, \lambda' v_a]=\lambda \lambda'
[v_a,v_a]=0$ and hence the spaces $M_a$ fulfill the requirements of
the last section and define stabilizer states. Further, set
$\zeta^{(a)}_b(\lambda v_a)
:=\zeta_b(\lambda):=\exp(i\frac{2\pi}p \Tr(b \lambda))$. Each
$\zeta^{(a)}_b$ is easily seen to be a character of $M_a$. Now define
the stabilizer states
\begin{equation}
	\mathcal{B}^{(a)}_b:= d^{-1} \sum_{\lambda} \zeta_b(\lambda)\, 
	w(\lambda v_a).
\end{equation}
We claim that these states constitute a set of MUBs. Indeed
\begin{eqnarray}
	\tr \mathcal{B}^{(a)}_b \mathcal{B}^{(a')}_{b'}
	&=&
	d^{-2} \sum_{\lambda \lambda'} \zeta_b(\lambda)
	\zeta_{b'}(\lambda')
	\tr w(\lambda v_a + \lambda' v_a')	
	\nonumber \\
	&=&
	d^{-1} \sum_{\lambda \lambda'} \zeta_b(\lambda)
	\zeta_{b'}(\lambda')
	\delta_{\lambda v_a,\lambda' v_a'}
	\label{eqn:mubLastLine}
\end{eqnarray}
Now if $a=a'$ then Eq.\ (\ref{eqn:mubLastLine}) reduces to
\begin{eqnarray}
	\tr \mathcal{B}^{(a)}_b \mathcal{B}^{(a')}_{b'}
	&=& d^{-1} \sum_{\lambda} \exp\big(i\frac{2\pi}p 
	\Tr(\lambda (b-b'))\big) \nonumber\\
	&=& \delta_{b,b'}
\end{eqnarray}
while for $a\neq a'$ we use the property of the finite plane $F^2$
that the two lines $M_a$ and $M_{a'}$ intersect exactly at $\lambda=0$
to conclude
\begin{eqnarray}
	\tr \mathcal{B}^{(a)}_b \mathcal{B}^{(a')}_{b'}
	&=& d^{-1} \zeta_b(0)\zeta_{b'}(0)=d^{-1}.
\end{eqnarray}
Hence, for a fixed $a$, the set $\{\mathcal{B}^{(a)}_b\}_b$ forms a
basis and all $d$ bases corresponding to different values of $a$ are
mutually unbiased. The computational basis corresponds to the set
$M_\infty=\{\lambda (1,0)^T\,|\,\lambda \in F\}$. Repeating the
reasoning employed above, one finds that it is unbiased with respect
to all the other ones; hence, we have constructed a maximal set of
$d+1$ MUBs.

\subsection{Asymptotic designs from MUBs}
\label{sec:approximations}

This section revolves around the following definition.

\begin{definition}[Asymptotic $2$-designs]
	Let $I\subset \NN$ be an index set.
	A family of sets of unitaries $\D_d$, $d\in I$ is an
	\emph{asymptotic 2-design} if the matrices in $\D_d$ are
	$d$-dimensional and
	\begin{equation}
		\lim_{d\to\infty}\mathcal{P}(\D_d)=2.
	\end{equation}	
\end{definition}

A priori, it is not clear that this definition has any physical
relevance. After all, it is conceivable that, even though the frame
potential of $\D_d$ converges, the $\D_d$-twirling channels $T_{\D_d}$
do not become close to the $UU$-twirling channel in any sensible
metric. Indeed, the question of whether asymptotic designs are
``almost as good'' as strict ones cannot be answered in general, but
depends on the application one has in mind. One particular aspect of
this question will be illuminated in Lemma \ref{lem:convergence}.  We
will show that the series of twirling channels $T_{\D_d}$ does
converge to $T_{UU}$ in $D_\text{pro}$-norm. The latter norm has been
defined in Ref.\ \cite{gilchrist}, a well-readable account of the
merits and perils of different metrics for quantum channels.
Specifically, let $\Lambda$ and $\Lambda'$ be channels with respective
Choi matrices $C, C'$.  If $\Delta=C-C'$, then
$D_\text{pro}(\Lambda,\Lambda'):=d^{-1} \tr|\Delta|$.  Some physical
interpretations of $D_\text{pro}$-convergence are listed in Ref.
\cite{gilchrist}.

\begin{lemma}\label{lem:convergence}
	Let $\D_d$ be an asymptotic 2-design. Then the $\D_d$-twirling
	channels $T_{\D_d}$ converge to $T_{UU}$ in $D_\text{pro}$-norm.
\end{lemma}

\begin{proof}
	Let $C_{\D_d}, C_{UU}$ be the usual Choi matrices, let
	$\Delta=C_{\D_d}-C_{UU}$. As quantum channels preserve Hermiticity,
	the Choi matrices and hence $\Delta$ are Hermitian. Let
	$\{\delta_i\}$ be the set of eigenvalues of $\Delta$.
	By Corollary \ref{cor:choiDistance}, we know that $\tr |\Delta|^2
	\to 0$. Hence, for all $i$, $|\delta_i|<1$ eventually and therefore,
	for large enough $d$:
	\begin{eqnarray*}
		D_\text{pro}(\Lambda_d,\Lambda_{UU})
		&=& d^{-1} \sum_i |\delta_i|  < d^{-1} \sum_i |\delta_i|^2 \\
		&=& d^{-1} ||\Delta||_2^2 \to 0.
	\end{eqnarray*}
\end{proof}

The technically non-trivial part of this section is contained in the
next theorem.

\begin{theorem}[Mutually unbiased bases]
\label{thm:mubs}
	Let $d=p^m$ be a power of a prime. 
	Then in $\H=\CC^d\otimes \CC^d$
	exist $d^2+1$ MUBs of which $d^2-d$ bases 
	are maximally entangled
	and $d+1$ bases factor.
\end{theorem}

\begin{proof}\emph{(of Theorem \ref{thm:mubs})}
	Let $B=\FF_d$, $F=\FF_{d^2}$. Using the notation of Section
	\ref{sec:weylAndCo}, we assume that a basis for $F$ over $B$ has been
	chosen. For $a,b\in F$, let $\mathcal{B}^{(a)}_b$ be a projection onto
	a stabilizer state in $\H$, as defined in Section
	\ref{sec:mubs}. Hence
	\begin{eqnarray*}
		\mathcal{B}^{(a)}_b 
		&=& d^{-2} \sum_{(p,q)\in M_a} \zeta^{(a)}_b(p,q) w_F(p,q)\\
		&=& d^{-2} \sum_{(p,q)\in M_a} \zeta^{(a)}_b(p,q) w_B(p_1,q^1)\otimes
		w_B(p_2,q^2).
	\end{eqnarray*}
	Defining $N_a=\{(p,q)\in M_a\,|\,p_2=q^2=0\}$, we get
	\begin{eqnarray}\label{eqn:mubReduction}
		\tr_2 \mathcal{B}^{(a)}_b 
		&=& d^{-1} \sum_{(p,q)\in N_a} \zeta^{(a)}_b(p,q) w_B(p_1,q^1).
	\end{eqnarray}
	Clearly, $N_a$ is a $B$-vector space, so it has cardinality
	$|N_a|=d^n$ for some $n$. If $n=0$, then 
	$\tr_2 \mathcal{B}^{(a)}_b$ equals $\Id_1$, so the state was
	maximally entangled. If $n=1$, Eq.\ (\ref{eqn:mubReduction}) is of
	the form of Eq.\ (\ref{eqn:generalStabilizer}) which makes
	$\tr_2 \mathcal{B}^{(a)}_b$ a pure state on $\CC^d$. 
	Further, $n=2$ would imply $|N_a|=|M_a|$ and hence
	$\mathcal{B}^{(a)}_b=\rho_1\otimes \Id_2$ for some density operator
	$\rho_1$, which is impossible as $\mathcal{B}^{(a)}_b$ is pure.
	Lastly, $n>2 \Rightarrow |N_a|>|M_a|$, which is absurd. Hence any
	vector in the standard set of MUBs is either a product or else
	maximally entangled. 

	Now, $(a \lambda, \lambda)\in N_a$ if and only if $\Tr_{F/B}(a
	\lambda e^2)=\Tr_{F/B}(\lambda e_2)=0\Leftrightarrow \lambda =
	\lambda_1  e^1 \wedge a \lambda = (a\lambda)^1 e_1$.
	Assume that 
	\begin{equation}\label{eqn:condOnA}
		a=b \frac{e_1}{e^1}
	\end{equation}
	for some $b\in B$. Then $\lambda=\lambda_1 e^1 \Rightarrow \lambda a
	= \lambda_1 b e_1$ and hence $|N_a|=d$. Conversely, assume that
	$|N_a|=d$. Then for all $\lambda=\lambda_1 e^1$ we must have that
	$\lambda a = b e_1$ for some $b\in B$. Solving for $a$ shows that Eq.
	(\ref{eqn:condOnA}) must hold.
	Hence among the $d^2$ bases associated with the sets $M_a$, there are
	exactly $|B|=d$ factoring ones. Taking the computational basis into
	account, the assertion becomes immediate.
\end{proof}

The validity of Theorem \ref{thm:mubs} implies the existence of
asymptotic 2-designs.

\begin{corollary}[Existence of asymptotic 2-designs]
	Let $I$ be the set of prime-power integers. Then there exists an
	asymptotic 2-design $\D_d$, for $d\in I$.
\end{corollary}

\begin{proof}
	We compute the frame potential of the $d^2(d^2-d)=d^4-d^3$
	unitaries $\D_d$ which can be constructed via Eq.\ (\ref{eqn:vec})
	from the maximally entangled MUB vectors of Theorem \ref{thm:mubs}:
	\begin{eqnarray*}
		\mathcal{P}(\D_d)
		&=& d^4 \sum_{a,a'} \sum_{b,b'} 
		|\braket \psi^{(a)}_b, \psi^{(a')}_{b'}.|^4 /K^2\\
		&=& d^4 \left( 1 + (d^2-d-1) d^{-2} \right) /K \\
		&=& \frac{2d^4-(d^{-1}+d^{-2})}{d^4-d^3} \to 2 \quad(d\to\infty).
	\end{eqnarray*}
\end{proof}

\subsection{Clifford designs}
\label{sec:cliffordDesigns}

Let us review the technique employed in
Ref.\ \cite{dataHiding,dankertThesis,dankert} to construct a 2-design.
The construction proceeds in two steps. First one realizes that
twirling an operator $\rho$ by Weyl matrices reduces $\rho$ to its
``Weyl-diagonal'' components (see below). Secondly, twirling the
resulting operator using the metaplectic unitaries $ \mu(S)$
``evens out'' the coefficients to yield a $U\otimes U$-invariant state.

Denote by $T_W$ the \emph{Weyl twirl channel}
\cite{dataHiding,dankertThesis}:
\begin{equation}
	T_W(\rho) = d^{-2} \sum_{a\in V} w(a)\otimes w(a)
	\,\rho\,w(a)^\dagger\otimes w(a)^\dagger.
\end{equation}
Expanding $\rho = \sum_{b, b'\in V} \rho_{b,b'}\,w(b)\otimes w(b')$, we
compute: 
\begin{eqnarray}\label{eqn:weyltwirl}
	&& T_W(\rho) \nonumber \\
	&&= d^{-2} \sum_{a,b,b'} \rho_{b,b'}\,w(a)w(b)w(a)^\dagger\otimes 
	w(a)w(b')w(a)^\dagger\nonumber  \\
	&&= d^{-2} \sum_{b,b'} \rho_{b,b'}\, w(b)\otimes w(b') 
	\sum_{a} \chi(-[b+b',a])\nonumber  \\
	&&= \sum_b \rho_{b,-b}\, w(b)\otimes w(-b). 
\end{eqnarray}
The final transformation follows from the fact that
$\chi([b,\,\cdot\,])$ is a non-trivial character of $V$ for any $b\neq
0$.  

The \emph{Clifford twirl channel} \cite{dataHiding,dankertThesis} is
given by
\begin{eqnarray}\label{eqn:cliffordtwirl1}
	&&T_\mathcal{C}(\rho) \\
	&&= |\mathcal{J}_{\FF_p^n}|^{-1} \sum_{U\in \mathcal{J}_{\FF_p^n}}
	(U\otimes U)\,\rho\,(U \otimes U)^\dagger \nonumber \\
	&&= |\mathcal{J}_{\FF_p^n}|^{-1} \sum_{S\in \Sp(p,n)} 
	 (\mu(S)\otimes \mu(S) )\,T_W(\rho)\,
	 (\mu(S) \otimes \mu(S))^\dagger, \nonumber
\end{eqnarray}
where $\Sp(p,n)$ denotes the group of symplectic matrices on
$V=\FF_p^{2n}$.
Using the notation and results of Eq.\ (\ref{eqn:weyltwirl}), one
concludes:
\begin{eqnarray}\label{eqn:cliffordtwirl2}
	&&T_C(\rho) \nonumber \\
	&&=|\Sp(p,n)|^{-1} \sum_{b\in  V} \rho_{b,-b}\,
	\sum_{S\in \Sp(p,n)}  w(S b) \otimes w(-S b) \nonumber \\
	&&=\alpha\,\Id + \beta\, \sum_{0\neq b \in  V} w(b)\otimes w(-b),
\end{eqnarray}
where the constants are given by
\begin{equation}
	\alpha = \rho_{0,0},\quad\quad
	\beta=\sum_{0\neq b \in  V} \rho_{b,-b}.
\end{equation}
Eq.\ (\ref{eqn:cliffordtwirl2}) follows because the symplectic
group acts transitively on $V^\sharp:=\{a\in V\,|\,a\neq 0\}$ and hence
maps every element of $V^\sharp$ equally often to every other element.
It is evident that $T_\mathcal{C}$ projects onto exactly two subspaces
and is hence equal to $T_{UU}$ by Section \ref{sec:groupDesigns}.

Now let $G$ be some subgroup of $\Sp(p,n)$ such that $G$ acts
transitively on $V^\sharp$. From the above argument it is clear that
$T_{\mu(G)}\circ T_W=T_{UU}$ and hence that $\{w(v)\mu(S)\,|\,v\in V,
S \in G\}$ is a unitary 2-design. An obvious choice is to set $G=\Sp(p^n,1)$
which is the basis of Ref.\ \cite{chau}. The advantage of going from
the multi-particle picture to the single-particle picture is an
exponential reduction of the cardinality of the design:
\begin{eqnarray}\label{eqn:cliffordCardinalities}
	|\Sp(p,n)| &=& p^{n^2} \prod_{i=0}^{n-1} (p^{2(n-i)}-1)  \\
	&=& O(p^{2n^2+n})=O(d^{2(\log_p d)^2 + \log_p d}),\nonumber\\
	\nonumber\\
	|\Sp(p^n,1)| &=& p^{n}(p^{2n}-1)  = O(p^{3n}) = O(d^3)
\end{eqnarray}
(see Ref.\ \cite{huppert} for a derivation). What possibilities are
there to further improve the cardinality? Clearly, any group acting
transitively on $V^\sharp$ must have order $k\,|V^\sharp|=k(d^2-1)$
for some integer $k$. The smallest value is $k=1$ and hence $d^2
(d^2-1)$ gives a lower bound to the number of elements a Clifford
design arising from such a construction can have (c.f. Conjecture
\ref{conj:cliffordBound}). In Ref.\ \cite{chau} Chau showed that for
$d=2,3,5,7,11$, such minimal subgroups do exist.  He goes on to rule
out the existence of any subgroup $G$ of $\Sp(p^n,1)$ which has
cardinality a multiple of $V^\sharp$. Hence no reduction below $d^2
|\Sp(d,1)| = d^5-d^3$ seems to be possible in general.

However, the argument leaves open the possibility of finding subgroups
$G$ of $\Sp(p,n)$ which act transitively on the non-zero elements of
the vector space and are smaller than $|\Sp(p^n,1)|$. While we do not
know if such groups exist in general, we know of one example for $d=9$.
In Table \ref{tbl:generators} we list the generators of a transitively
acting subgroup $G$ of $\Sp(3,2)$ of order $160=2 |V^\sharp|$.  It
yields a Clifford 2-design of cardinality $2(d^4-d^2)=12,960$, where
the design induced by $\Sp(9,1)$ has $58,230$ elements and the one
associated with $\Sp(3,2)$ consists of $4,199,040$ unitaries. All claims
made about the generators can easily be tested by a computer algebra
system.

\section{Miscellaneous topics}

\subsection{Higher orders and general groups}
\label{sec:higherOrder}

We saw in Section \ref{sec:generalPreliminaries} that the effect of
the twirling channel $T$ induced by some group $G$ and a corresponding
unitary representation $U_g$ depends only on the the decomposition of
$U_g$ into irreducible constituents (see Appendix
\ref{sec:generalTwirling} for a version of Eq.
(\ref{eqn:twirlingProjections}) valid for general representations).

Based on this observation, the notion of a group design can easily be
generalized: 

\begin{definition}[General group designs]
Let $\mathcal{U}$ be some group of unitary matrices.
Then a finite subgroup $\mathcal{D}$ of $\mathcal{U}$ is a
\emph{$\mathcal{U}$-group design of order $t$} if the $t$-th tensor
power of $\mathcal{D}$ has the same number of irreducible constituents
as the $t$-th tensor power of $\mathcal{U}$.
\end{definition}

The most natural setting for applying the above definition is given by
unitary designs $\mathcal{U}=U(d)$ of higher order $t>2$. How many
irreducible constituents do we expect the representation $U\mapsto
U^{\otimes t}$ to decompose into? The following lemma answers this
question by giving the frame potential for unitary $t$-designs, at
least in two special cases.

\begin{lemma}\label{lm:tPotential}
	The frame potential of a unitary $t$-design in dimension $d$ is
	given by
	\begin{equation*}
		\begin{array}{cl}
			t!& \text{for }d\geq t,\\
			\\
			\sum_{i=0}^{\lfloor n/2 \rfloor} 
			\big(\frac{n! (n-2i+1)}{i! (n-i+1)}\big)^2
			\quad& \text{for }d=2.
		\end{array}
	\end{equation*}
\end{lemma}

Once again, we can search the GAP library for examples. Table
\ref{tbl:5design} gives an example of a matrix group $G$ in $d=2$,
whose 5th tensor power decomposes into 42 irreps, the required value
for a 5-design.  As a matter of fact, $G$ is very close to being a
6-design: its 6th tensor power has 133 irreducible components, whereas
the full unitary group $U(2)$ decomposes into only 132 irreps.  Note
that the matrices in Table \ref{tbl:5design} are not unitary in the
standard basis. However, as is well-known \cite{lieRepTheory}, any
representation of a finite group is equivalent to a unitary one:
one can easily construct a similarity transformation mapping the
given matrices to unitaries.

\begin{proof} \emph{(of Lemma \ref{lm:tPotential})}
	The following facts are well-known \cite{lieRepTheory}: (i) there is a
	one-one correspondence between irreducible components of 
	the $t$-th
	tensor power of $U(d)$ and young frames $\mathcal{F}$ partitioning
	the integer $t$ into no more than $d$ parts; (ii) the multiplicity
	of the irrep belonging to a specific frame $\mathcal{F}$ is given
	by the dimension $d_\mathcal{F}$ of the corresponding irrep of $S_t$.

	If $d\geq t$, the restriction ``no more than $d$ parts'' becomes
	irrelevant. Using the results of Section \ref{sec:characters}, we find
	that the frame potential of a $t$-design is
	\begin{equation*}
		\sum_{\mathcal{F}} d_\mathcal{F}^2=|S_t|=t!,
	\end{equation*}
	where the sum is over all young frames with $t$ boxes. 
	The second claim follows in a similar fashion, using well-known
	formulas for $d_\mathcal{F}$ (which can be found, e.g., in Ref.\ 
	\cite{lieRepTheory}).
\end{proof}

\subsection{Energy-preserving operations as in linear optics}

We will briefly comment on a specific representation of the unitary
group, which plays a prominent role in linear optics. General
references for the introductory paragraphs are Refs.
\cite{GBook,Symplectic,LO,CV}. The most central 
mathematical object in
the description of physical systems of $d$ bosonic modes are the
creation and anihilation operators $a_k, a_k^\dagger$. Recall that the
set of unitaries on $L^2(\RR)$ which 
keep the vector space spanned by
the $a_k, a_k^\dagger$ invariant under conjugation 
form a projective
representation of the real symplectic group $Sp(2d)$. 
The representation is
referred to as the \emph{metaplectic representation}. Bosonic systems
are often thought about in terms of their \emph{phase space
description}, where these metaplectic operations appear in an
especially natural way.

The maximal compact subgroup of $Sp(2d)$ is 
given by the intersection
of $Sp(2d)$ with the set of orthogonal transformations
$SpO(2d):=Sp(2d)\cap O(2d)$.  
Physically, the elements of $SpO(2d)$
correspond to energy-preserving or {\em passive operations}. These
operations are exactly the ones which are easily accessible in the
laboratory using passive linear optical elements (phase
shifts and beam splitters, but no squeezers, which 
correspond to elements in $Sp(2d)$ which are not contained
in $O(2d)$).
It is a well-known fact that $SpO(2d)\simeq U(d)$, which might 
trigger some hope that our theory could be applicable to 
these systems.
However, the metaplectic representation is infinite-dimensional and in
this work we did not develop the means to cope with such
representations. 

In an indirect approach, we can, however,
nevertheless exploit the developed formalism:
Fortunately, important properties of bosonic quantum states can be
described entirely in terms of objects on the $2d$-dimensional phase
space. Indeed, define the \emph{canonical coordinates}
or quadrature operators 
$r= (x_1,\dots,x_d,p_1,\dots, p_d)$
by 
\begin{equation}
	x_k:=(a_k+a_k^\dagger)2^{-1/2},\,\,\,
	p_k:=i (a_k^\dagger- a_k)2^{-1/2}.
\end{equation}	 
A much-studied object in particular in 
quantum optics are the various second moments of the quadrature
operators with respect to a given state. 
Assuming that all first moments vanish, 
the second moments can be 
conveniently assembled in a real symmetric 
$2d\times 2d$ \emph{covariance matrix} $\gamma$, 
defined as
\begin{equation}
	{\gamma^k}_{l} = 2 \left(\tr(r_k r_l)+ \tr(r_l r_k)	
	\right).
\end{equation}
An interaction process preserving the energy
would then give rise to a map
\begin{equation}
	\gamma\mapsto 
	S\gamma S^T = :\gamma',
\end{equation}
where $S\in SpO(2d)$. The following proposition assures that unitary
designs can be used to tackle problems in this context.

\begin{proposition}[Averaging over passive operations]
	\label{prop:spo}
	Let $\mathcal{D}\subset U(d)$ be a unitary group design of order $t$. 
	The image of $\mathcal{D}$ in $SpO(2d)$ under the usual isomorphism
	is then a $SpO(2d)$-group design of order $t$.
\end{proposition}

	A setting that can be studied using designs in this way is the
	following: consider a system of $d$ interacting bosons. Having a
	``microcanonical ensemble'' in mind, we might be interested in the
	expected value of various quantities after random energy-preserving
	interactions have been applied. Haar averages 
	over $SpO(d)$ would
	constitute a sensible model for such a system (see also
	Ref.\ \cite{alessio}).
	
Let us give a concrete example. Take $\{|i\rangle\}_i$ as a basis of 
the vector complex space in which the complex moment
matrices are defined. It is straightforward to see that 
the mean energy of the first mode is given by
$E= \bra 1 . \Omega \gamma \Omega^\dagger \ket 1.$.
The quantity
	\begin{eqnarray}
		\Delta E & =& 
		\int_{U(n)} 
		\bra 1 .
		(U\oplus \bar U) \Omega \gamma \Omega^\dagger 
		(U\oplus \bar U)^\dagger \ket 1. ^2 
		dU\\
		&-&\biggr(
		\int_{U(n)} 
		\bra 1 .
		(U\oplus \bar U) \Omega \gamma \Omega^\dagger 
		(U\oplus \bar U)^\dagger \ket 1. 
		dU\biggr)^2 . \nonumber
	\end{eqnarray}
	gives the \emph{expected energy fluctuations} 
	of the first mode and is
	directly amenable to evaluation using unitary 2-designs.

\begin{proof}[Proof (of Proposition \ref{prop:spo}).]
	The usual isomorphism between elements $U\in U(d)$ and
	elements $S\in SpO(2d)$ can be stated as follows: If
	$	U = X + i Y$,
	then $S$ is given by \cite{Symplectic,Passive}
	\begin{equation}
	S(U)= \left(
	\begin{array}{cc}
	X & Y\\
	-Y & X
	\end{array}
	\right).
	\end{equation}
	Setting
	\begin{equation}
			\Omega =
			\left(
		\begin{array}{cc}
		\Id & i \Id\\
		\Id & -i \Id
		\end{array}
		\right)/\sqrt{2},
	\end{equation}
	one finds easily that
	\begin{equation}
		\Omega S(U) \Omega^{-1} =	U\oplus \bar U.
	\end{equation}
	As $\bar{\mathcal{D}}$ is certainly a group $t$-design if
	$\mathcal{D}$ is, the statement follows.
\end{proof}

\subsection{Random entanglement}

Originally posed by Lubkin and later popularized by Page, the
following questions has a long history: what is the average
entanglement of a composite system in a pure state
\cite{Page,Generic,Oscar,alessio}. One motivation for studying
such problems is to justify the ad hoc ``rule of minimal prejudice''
employed in statistical physics, which states that among the ensembles
compatible with macroscopic observables the one maximizing the entropy
is realized in nature. The problem was originally stated in terms of
the entropy of entanglement of subsystem A: $S(\rho_A) = -\tr(\rho_A
\log \rho_A)$, but various other measures, for example the purity
$\tr(\rho_A^2)=||\rho_A||_2^2$ of $\rho_A$ can be used. The latter
quantity has the advantage that its expectation value
\begin{eqnarray}\label{eqn:pagesIntegral}
	\int_{U(d)}
	\| \,
	\tr_B[U\rho U^\dagger]\,
	\|_2^2 dU
\end{eqnarray}
is a Haar integral of a second-order polynomial and can thus be
directly evaluated by averaging over a 2-design (above, $\rho$
projects onto an arbitrary 2-system state vector $\ket\psi.$).

Based on this observation, we obtain a simple answer to a special
case of this problem as a corollary to Theorem \ref{thm:mubs} (c.f.\
Ref.\ \cite{zyk} for a much more general, but much longer proof).

\begin{corollary}[Average entanglement]\label{cor:page}
	Let $d$ be the power of a prime. The average entanglement of pure
	states on $\CC^d\otimes \CC^d$, as measured by the purity, is
	$2d/(d^2+1)$. 
\end{corollary}

\begin{proof}
	We choose $\rho=\ket0. \bra0. \otimes \ket0. \bra 0.$ 
	and average over the Clifford group
	$\mathcal{J}_{{d^2},1}$. The image of $\ket0.\otimes \ket 0.$ 
	under the action of
	$\mathcal{J}_{{d^2},1}$ constitutes of the bases of Theorem
	\ref{thm:mubs}. The purity of a reduced density matrix of a product
	state equals 1, for a maximally entangled state it is $d^{-1}$. We
	can thus compute the average by counting:
	\begin{eqnarray*}
		\frac{(d^2-d) d^{-1} + (d+1)}{d^2+1} = \frac{2d}{d^2+1}
	\end{eqnarray*}
	as claimed.
\end{proof}

Note that, while the question of finding the expected entanglement of
a pure state is stated in terms of analysis, we could answer this
special case  by purely combinatorial means.

\section{Summary}

In this work, we presented a first systematic analysis of the
mathematical structure of unitary designs. We pointed out a connection
to group representation theory, gave bounds on the number of elements
of a design, made the relationship to spherical designs explicit,
and used this connection to construct approximate unitary designs.
Foremost, the pivotal concept of a frame potential has been explored.
Intriguingly, the latter quantity appears very naturally in different,
seemingly unrelated areas.

While much insight into the structure of unitary designs has been
gained, many questions remain unresolved as 
interesting open problems: finding a systematic way for
constructing designs for any choice of parameters $t,d$ or improving
the bounds for their cardinalities, to name just two.

\section{Acknowledgments}

The authors thank O.\ Dahlsten, M.\ M{\"u}ller, 
and A.\ Serafini for
helpful discussions. 
Support has been provided by the DFG (SPP 1116),
the EU (QAP), the EPSRC, the QIP-IRC, Microsoft Research, 
and the
EURYI Award Scheme.

\section{Appendix}

\subsection{General twirling channels}
\label{sec:generalTwirling} 

Eq.\ (\ref{eqn:twirlingProjections}) gives an explicit formula for
a channel twirling over an \emph{irreducible} representation $U_g$. In
this section, we state the relation for the general case.  So assume
that $g\mapsto U_g$ decomposes into a set of irreps $U^{(i)}$, which
have dimension $d_i$ and occur with multiplicity $n_i$ respectively.
The underlying Hilbert space $\H$ then decomposes as
\begin{equation}
	\H=\bigoplus_i \H_i\otimes \CC^{n_i},
\end{equation}
where $\H_i=\CC^{d_i}$ (see, e.g., Ref.\ \cite{lieRepTheory}). The
representation $U_g$ itself can be written as
\begin{equation}
	U_g = \sum_i U^{(i)}_g \otimes \Id_{n_i}.
\end{equation}
Now set $P_i=\Id_{\H_i}\otimes\Id_{n_i}$.  The twirling channel
becomes
\begin{equation}
	T_A(\rho) = \sum_i \tr_{\H_i}(P_i' \rho) P_i,
\end{equation}
as can be checked without difficulty.

\newpage

\begin{table*}
\caption{Some group designs found by the GAP system. The group name
and character number refer to the names used by the ``ctllib''
package \cite{ctllib}.}
\label{tbl:gap}
\begin{tabular}{r|r|r|r|r}
	$d$			& $K$		& $K/(d^4-d^2)$ & Group		& Irred.\ character no. \\
	\hline
	2			& 12		& 1						& \verb 7^2:(3x2A4) & 10 \\
	3			& 72		& 1						& \verb 2^3.L3(2)   & 2  \\
	4			& 1920	& 8						& \verb 2.HSM10 		& 29 \\
	5			& 25920	& 43.2				& \verb 2^6:U4(2) 		& 2 \\
	6		  & 40320 & 32					& \verb 6.L3(4).2_1 	& 49\\
	8			& 20160 & 5						& \verb 4_1.L3(4) 		& 19 \\ 
	9			& 19440 & 3						& \verb 3.3^(1+4):2S5  & 25 \\
	10		& 190080 & 19.2				& \verb 2.M12.2 				& 22 \\
	11		& 13685760 & 942.			& \verb 6xU5(2) 				& 3 \\
	12		& 448345497600 & 21772800 & \verb 6.Suz 			& 153 \\
	13 		& 4585351680   	& 161501.  &\verb 2.S6(3) 		& 2 \\
	14 		& 87360					&  2.29		 &\verb Sz(8).3 		& 4 \\
	18		& 50232960			&	480			 &\verb 3.J3 				& 22 \\
	21		& 9196830720		& 47397.		&\verb 3.U6(2) 		& 47 \\
	26		& 17971200			& 39.				&\verb 2F4(2) 		& 2 \\
	28		& 145926144000	& 237714.		&\verb 2.Ru 			& 37 \\
	41		& 65784756654489600 & 23294225607. &\verb S8(3)  & 2 \\
	45		& 10200960					& 2.49		&\verb M23 	& 3 \\
	342		& 460815505920			& 34.			&\verb 3.0N  & 31 \\
	1333	& 86775571046077562880 & 27483822. &\verb J4 	& 2
\end{tabular}
\end{table*}

\begin{table*}
\caption{Generators of a subgroup $G$ of $\Sp(3,2)$ of order
$2(d^2-1)=160$. The group $G$ acts transitively on the non-zero
elements of the phase space $V=\FF_3^{4}$ and induces a unitary
design, as described in Section \ref{sec:cliffordDesigns}.}
\label{tbl:generators}
\begin{eqnarray*}
\left(
\begin{array}{cccc}
2 & 2 & 2 & 0 \\ 
1 & 2 & 2 & 0 \\ 
1 & 2 & 0 & 2 \\ 
0 & 0 & 1 & 1 \\ 
\end{array}
\right),
\left(
\begin{array}{cccc}
0 & 2 & 1 & 0 \\ 
0 & 0 & 0 & 2 \\ 
1 & 0 & 0 & 2 \\ 
0 & 2 & 0 & 0 \\ 
\end{array}
\right),
\left(
\begin{array}{cccc}
0 & 2 & 0 & 0 \\ 
2 & 0 & 0 & 0 \\ 
2 & 0 & 0 & 2 \\ 
0 & 1 & 2 & 0 \\ 
\end{array}
\right),
\left(
\begin{array}{cccc}
0 & 1 & 1 & 0 \\ 
1 & 0 & 0 & 2 \\ 
0 & 0 & 0 & 1 \\ 
0 & 0 & 1 & 0 \\ 
\end{array}
\right),
\left(
\begin{array}{cccc}
1 & 0 & 0 & 0 \\ 
0 & 2 & 0 & 0 \\ 
0 & 0 & 2 & 0 \\ 
0 & 0 & 0 & 1 \\ 
\end{array}
\right),
\left(
\begin{array}{cccc}
2 & 0 & 0 & 0 \\ 
0 & 2 & 0 & 0 \\ 
0 & 0 & 2 & 0 \\ 
0 & 0 & 0 & 2 \\ 
\end{array}
\right)
\end{eqnarray*}
\end{table*}

\begin{table*}
\caption{
	Generators of a matrix group $G$ of order 120. Up to a similarity
	transformation, the group gives rise to a unitary 5-design with 60
	elements. In fact, $G$ is an irreducible representation of
	$SL(2,\FF_5)$ affording the character $X.3$ as listed in the GAP
	character library \cite{ctllib}. The repersentation has been
	constructed using the ``Repsn'' package \cite{repsn}. Below,
	$\omega=\exp(\frac{2\pi}{15} i)$.
}
\label{tbl:5design}
\begin{eqnarray*}
\left(
\begin{array}{cc}
-1 & 0 \\
0  &-1 \\
\end{array}
\right),
\left(
\begin{array}{cc}
-\omega^{11}-\omega^{14} & -\omega^{11}-\omega^{14} \\
\omega^{10} & -\omega-\omega^4 
\end{array}
\right),
\left(
\begin{array}{cc}
-\omega-\omega^2-\omega^4-\omega^8-2(\omega^{11}-\omega^{14})& \omega^6+\omega^9 \\
\omega^{11}+\omega^{14} & -\omega^5
\end{array}
\right)
\end{eqnarray*}
\end{table*}


\begin{thebibliography}{99}
\bibitem{clifford}
	Be aware that the Clifford group which appears in the context of
	quantum information theory \cite{gottesman}
	has no connection to the Clifford group used e.g. in the
	representation theory of $SO(n)$ \cite{lieRepTheory}.
\end{thebibliography}
\end{document}